\documentclass[epjST]{svjour}
\usepackage[numbers, sort, sort&compress]{natbib}
\usepackage{graphics}
\begin{document}
\title{Chimera states in ensembles of excitable FitzHugh--Nagumo systems}
\author{Nadezhda Semenova\inst{1,2}\fnmsep\thanks{\email{nadya.i.semenova@gmail.com}}}
\institute{D\'{e}partement d'Optique P. M. Duffieux, Institut FEMTO-ST,  Universit\'e Bourgogne-Franche-Comt\'e CNRS UMR 6174, Besan\c{c}on, France. \and Department of Physics, Saratov State University, Astrakhanskaya str. 83, 410012 Saratov, Russia.}

\abstract{
An ensemble of nonlocally coupled excitable FitzHugh--Nagumo systems is studied. In the presence of noise the explored system can exhibit a special kind of chimera states called coherence-resonance chimera. As previously thought, noise plays principal role in forming these structures. It is shown in the present paper that these regimes appear because of the specific coupling between the elements. The action of coupling involve a spatial wave regime, which occurs in ensemble of excitable nodes even if the noise is switched off. In addition, a new chimera state is obtained in an excitable regime. It is shown that the noise makes this chimera more stable near an Andronov--Hopf bifurcation.
} %end of abstract
\maketitle
\section{Introduction}
\label{sec:intro}

All real systems are inevitably affected by noise. The impact of noise plays a principal role and strongly dictates the properties of the oscillatory dynamics. Noise influences lead to deterioration of the observed effects or to their complete destruction. However, this does not always lead only to destructive effects. It is known that noise can qualitatively change the oscillatory behaviour, and induce bifurcation phenomena which give rise to the appearance of new regimes of functioning \cite{HU93a,PIK97,NEI97,LIN04,USH05,ZAK10a,ZAK13}. Moreover, such transitions can be accompanied by increasing of regularity with growth of the noise intensity. Such cases may include, for example, coherence or stochastic resonance \cite{PIK97, BEN81}. However, the affected system must have some special characteristics and properties to demonstrate new effect in the presence of noise.

Noise induced effects in single oscillators are already quite well understood. Currently, most of open questions address the noise impact in networks and ensembles \cite{MAS17, YAM18, BUK18, MAJ19}. Ensembles of identical and non-identical oscillators initially have a wide range of possible regimes. Thus, apart from typical synchronization, desynchronization, spatial incoherence or coherence, such systems can simultaneously demonstrate the coexistence of several modes. An example of such coexistence are chimera states. It is a spatio-temporal pattern consisting of the neighboring clusters of elements with coherent and incoherent dynamics in one network. Initially, this effect was discovered by Kuramoto and Battogtoh \cite{KUR02a}, and then revealed in more details by Strogatz and Abrams \cite{ABR04, ABR08}. Since then, quite a long time has passed, and now this effect has already been found in the ensembles of many systems of various nature: phase oscillators, chaotic mappings, bistable elements, neuronal models, and many others \cite{MAJ19}. Usually one of the main conditions imposed on the ensemble is non-local connection, when each partial element has a finite radius of influence. Nowadays, chimera states have already been found in networks of various topologies \cite{PAN15, MAI15, BER17}, as well as in multilayer networks \cite{MAJ17, OME19}

The ensemble of nonlocally coupled neuron models is considered in the present paper. The FitzHugh--Nagumo system in excitable mode is chosen as a partial element. Initially, chimera states have already been found for an ensemble of similar systems but in oscillatory regime \cite{OME13}. Later, it has been shown that chimeras can be detected in the excitable mode in the presence of noise \cite{SEM16}. This effect was called coherence-resonance chimeras (CR chimera) because it combined the properties of chimeras and coherence resonance and was characterized by periodic switching of the coherent and incoherent parts position. Another feature was that CR chimera appeared only when the noise intensity belonged to a certain interval. If it was smaller, there were no oscillations in the system, but too strong noise led to complete spatial incoherence and desynchronization. 

This work is devoted to the study of coupling features being principal for realization of chimera states in ensembles of excitable oscillators. The comparative analysis of noise role and coupling peculiarities is carried out for the example of an ensemble of nonlocally coupled FitzHugh--Nagumo systems. In order to find out which coupling features lead to such a vulnerability of the system to noise, and whether other spatio-temporal regimes can be obtained taking these features into account.

\section{System under study}
\label{sec:system}
In the present paper the dynamics of a one-dimensional ensemble of nonlocally coupled FitzHugh--Nagumo systems is studied numerically. The ensemble is described by the following system of equations:
\begin{equation}\label{eq:ring_fhn}
\begin{array}{c}
\varepsilon\frac{du_i}{dt}=u_i-\frac{u^3_i}{3}-v_i +\frac{\sigma}{2R}\sum\limits_{j=i-P}^{i+P} [b_{uu}(u_j-u_i)+b_{uv}(v_j-v_i)], \\
\frac{dv_i}{dt}=u_i+a+ \frac{\sigma}{2R}\sum\limits_{j=i-R}^{i+R} [b_{vu}(u_j-u_i)+b_{vv}(v_j-v_i)] + \sqrt{2D} \xi_{i}(t),
\end{array}
\end{equation}
where $u_i$ and $v_i$ are activator and inhibitor variables of $i$th oscillator. The number of all nodes is $N$. The dynamics of partial elements depends on two parameters: $\varepsilon$ and $a$. The first one controls the time scale in the system and $a$ is the threshold parameter. Depending on its values, the individual FHN element can demonstrate an oscillatory ($|a|<1$) or excitable ($a$>1) regime. In oscillatory regime the system exhibits the periodic behaviour associated with the limit cycle in the phase plane. In the excitable regime all oscillations decay and there is only one stable fixed point in the phase plane. But this regime is interesting because of a coherence resonance. This effect consists in stochastically excited spiking, which is the most regular at a certain noise intensity range \cite{PIK97}. The system (\ref{eq:ring_fhn}) contains white Gaussian noise source $\xi_i(t) \in R$ of the intensity $D$. Index $i$ indicates that the noise sources are not correlated inside the network, and each oscillator has only its own noise source. All of them have the same statistical characteristics and intensities.

The components of Eq.~\ref{eq:ring_fhn} with sums describe the connectivity. Here a parameter $\sigma$ is the coupling strength. $P$ is the number of elements connected with each $i$th oscillator on the either side. Normalizing this value by the total number of elements, we get the coupling radius $r=P/N$. Under the sign of the sum there are two summands in each of the equations. In the first equation the parameter $b_{uu}$ defines the contribution of $u$-variables of the connected neighbours. A coefficient $b_{uv}$ is the same for the variables $v$ in the first equation. Parameters $b_{vu}$ and $b_{vv}$ in the second equation have the same meaning. Thus, the parameters $b_{uv}$ and $b_{vu}$ are responsible for the cross-linking. The type of connection in Eq.~\ref{eq:ring_fhn} came from neuroscience. It is described in more detail in \cite{KOZ98, VAN04a, HEN11}. Taking into account the large number of parameters, it is reasonable to enter one main operator controlling all four $b$-parameters. To do this, the rotational coupling matrix is introduced:

\begin{equation}
B = \left(
\begin{array}{ccc}
b_{\mathrm{uu}} & & b_{\mathrm{uv}} \\
b_{\mathrm{vu}} & & b_{\mathrm{vv}}
\end{array}
\right) =
\left(
\begin{array}{ccc}
\cos \phi  & & \sin \phi \\
-\sin \phi  & & \cos \phi
\end{array}
\right).
\label{eq:matrix_B}
\end{equation}

Now there is only one parameter $\phi$, which controls the impact of cross-linking ($b_{uv}$, $b_{vu}$) and self-linking ($b_{uu}$,  $b_{vv}$) in the equation.

\section{Coherence-resonance chimera}\label{sec:CR_chimera}

Chimera states in the ensembles with nonlocal coupling have been found for the same values of $\phi=\pi/2-0.1$ in both oscillatory and excitable regime \cite{OME13, SEM16}. In this section the coupling parameters $\sigma=0.4$ and $r=0.2$ are fixed for coherence-resonance chimera (CR chimera). These parameters correspond to CR chimera with one incoherent and coherent domains. With smaller values of the coupling radius the number of these clusters increases. So, for $r=0.12$ there are two incoherent domains in the network, and at $r=0.08$ there are three of them. The chimera state exists in a quite wide range of $\sigma$ parameter values, but at $\sigma<0.2$ there is a transition to spatial incoherence \cite{SEM16}. 

\begin{figure}
\begin{center}
	\resizebox{0.75\columnwidth}{!}{%
  	\includegraphics{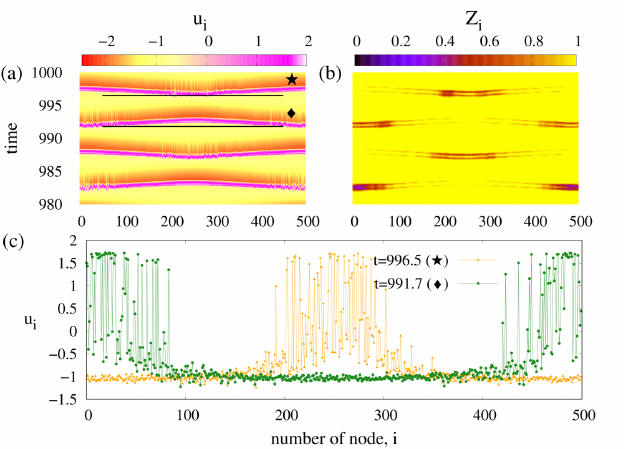} }
\end{center}
\caption{Coherence resonance chimera with corresponding space-time plot for the variable $u^t_i$ (a), local order parameter $Z^t_i$ (b) and snapshots (c) at time $t=996.5$ (orange curve) and $t=991.7$ (green curve). Parameters: $\phi=\pi/2-0.1$, $\varepsilon=0.05$, $a=1.001$, $r=0.2$, $\sigma=0.4$, $D=0.0002$.} 
\label{fig:CR_chimera}      
\end{figure}

Figure \ref{fig:CR_chimera},a shows the typical for CR chimeras space-time plot. Periodic switching the positions of incoherent areas makes it impossible to calculate statistical characteristics that involve a time realization. Therefore, the optimal characteristic for this chimera is the local order parameter \cite{WOL11a}, shown in Fig. \ref{fig:CR_chimera},b. It indicates the existence of coherent and incoherent domains at one time moment and does not require the temporal evolution. The local order parameter can be obtained as follows: 

\begin{equation} \label{eq:local_order_parameter}
 Z_k=\Big|\frac{1}{2\delta_Z}\sum\limits_{|j-k|\leq\delta_Z} e^{i \Theta_j}\Big|, \ \ \ k=1,\dots N
\end{equation}
where the geometric phase of the $j$th element is defined by $\Theta_j=arctan(v_j/u_j)$ \cite{OME13}. The values $Z_k = 1$ and $Z_k<1$ indicate coherent and incoherent domain, respectively. In this paper the parameter $\delta_Z$ is fixed $\delta_Z=20$.

%%%%%%%%%%%%%%%%%%%%%%%
%%%%%%%%%%%%%%%%%%%%%%%
%%%%%%%%%%%%%%%%%%%%%%%
%%%%%%%%%%%%%%%%%%%%%%%

Figure \ref{fig:CR_chimera},c shows the instantaneous spatial profiles (snapshots) of wave profiles in different half-periods. This panel clearly shows that the position of the incoherent part is first located on the edges of the ensemble, and then switches to the middle. Despite the seeming instability of the state, it is saved even at $t=10^6$.

It was obtained that for the value of $\phi=\pi/2-0.1$ CR chimera can be found only near an Andronov--Hopf bifurcation, at $0.995\le a\le 1.004$. The noise should have an intensity of $D\in[0.000062; 0.000325]$. Too large noise impact causes spatial incoherence. If the noise intensity is less than this interval, the ensemble does not show fluctuations. All deviations and initial conditions lead phase trajectories to a stable fixed point. On this assumption, it has been assumed earlier that coupling peculiarities at $\phi=\pi/2 -0.1$ lead to a special position of nullclines that makes system more sensitive to noise influence. The incoherent part is incoherent due to the fact that its nodes demonstrates spikes induced exclusively by noise. The rest of the nodes, on the other hand, are spiking only because of the coupling \cite{ZAK17}. The nature of the switching effect has not been revealed.

This work is dedicated to finding the reasons of CR chimeras. It seems that the property of these switches should also be caused by the specifics of coupling. However, in order to ensure that the effect of the coupling is not confused with the effect of noise, the latter must be excluded. To do this, set the noise intensity to $D=0$. In this case the same parameters as was for CR chimeras, lead to the complete disappearance of any oscillations. The noise influence can often lead to a shift of bifurcation values \cite{MEU88,USH05}. Therefore, a small variation of the parameter $\phi$ may lead to the CR chimeras predecessor. And this regime is found for $\phi=\pi/2+0.1$. 

\begin{figure}
\begin{center}
	\resizebox{0.75\columnwidth}{!}{%
  	\includegraphics{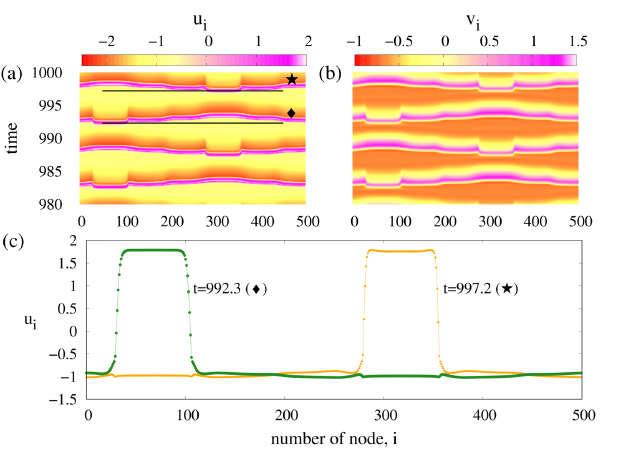} }
\end{center}
\caption{Switching spatial wave with corresponding space-time plot for the variables $u^t_i$ (a) and $v^t_i$  (b); and snapshots (c) at time moments $t=997.2$ (orange curve) and $t=992.3$ (green curve). Parameters: $\phi=\pi/2+0.1$, $\varepsilon=0.05$, $a=1.001$, $r=0.2$, $\sigma=0.4$, $D=0$.} 
\label{fig:CR_chimera_parent}       % Give a unique label
\end{figure}

Figure \ref{fig:CR_chimera_parent} shows the spatio-temporal profile of the new regime. On the snapshots (Fig. \ref{fig:CR_chimera_parent},b) some ``stair'' is clearly visible, and its position switches in time. If it is a stable regime existing without noise, the question arises what is the reason of these switching effect and what happens to nullclines at the same time?

\begin{figure}
\begin{center}
	\resizebox{0.75\columnwidth}{!}{%
  	\includegraphics{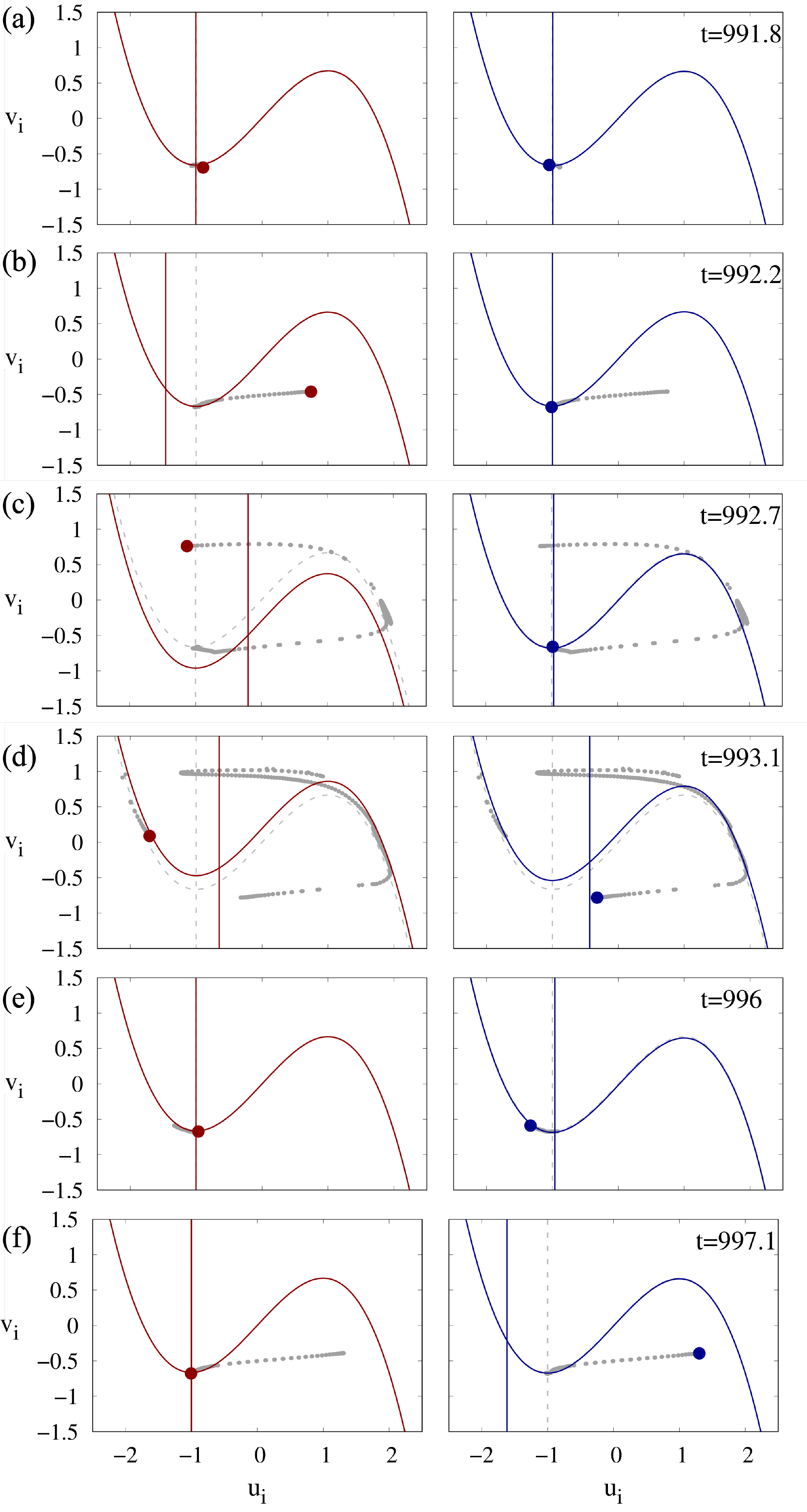} }
\end{center}
\caption{Nullclines (\ref{eq:nullclines}) for spatial wave shown in Fig.~\ref{fig:CR_chimera_parent}. The red points indicate the node $i=60$ and its nullclines. Blue color corresponds to the oscillator $i=315$ and its nullclines. The other are shown by dark-gray color. All points shows the projections of oscillators on the phase plane $\dot{u}$, $\dot{v}$. Gray lines represent nullclines for one isolated FitzHugh--Nagumo system under the same values of parameters $\varepsilon=0.05$, $a=1.001$. Other parameters: $\phi=\pi/2+0.1$, $r=0.2$, $\sigma=0.4$, $D=0$.} 
\label{fig:parent_nullclines}       % Give a unique label
\end{figure}

The location of the nullclines and information on their intersection provides the information on the location of the equilibrium states. The nullclines for one isolated FHN system can be obtained by solving a system of equations $\dot{u}=0$, $\dot{v}=0$. It leads to the solution: $u=const=-a$, $v(u)=u-u^3/3$. So, if $a=1.001$ there is a nullcline crossing at the point $u_0=-1.001$ and $v_0=u_0-u^3_0/3$. These values correspond to the coordinates of the equilibrium state in one isolated FitzHugh--Nagumo system. However, the presence of a stable wave regime (Fig.~\ref{fig:CR_chimera_parent}) in the ensemble at the same values of parameters suggests that the equation should include the influence of the coupling. Taking into account the components of the coupling, these equations are:
\begin{equation}\label{eq:nullclines}
\begin{array}{c}
v(u)=u-u^3/3 + C_u \\
u = -a - C_v,
\end{array}
\end{equation}
where $C_u$, $C_v$ are additional values produced by the nonlocal coupling. By converting the equation (\ref{eq:nullclines}) onto the ensemble, we can consider what happens to the nullclines of the nodes belonging to two different stairs of observed regime (Fig.~\ref{fig:CR_chimera_parent},c). The corresponding nullclines and projections to the phase plane are shown in Fig.~\ref{fig:parent_nullclines}.

The gray lines in Fig.~\ref{fig:parent_nullclines} represent nullclines for one isolated FHN system. The colors show the nullclines taking into account the connections (\ref{eq:nullclines}) for the oscillators $i=60$ (red) and $i=315$ (blue). Panel (a) shows that changes of nullclines for the first spiking oscillator are not essential. Initially, when the chimera states were considered, it was predicted that the incoherent part of the chimera started to spike only because of the noise, and the oscillators from the coherent domain make the path along the cycle only because of the coupling. That is why last oscillators make the spiking behaviour more coherent. Here the red oscillator $i=60$ should start to move along the cycle first, but there is almost no change in its nullclines. Moreover, now there is no noise which could excite the oscillations. Nevertheless, it starts to move along the limit cycle (b). The blue oscillator belonging to the opposite stair is still near the equilibrium state. Significant changes in nullclines occur only when most of the oscillators are on the opposite side of the limit cycle (c). When all the oscillators have already begun their journey through the cycle, the blue oscillator joins them. The panel (d) clearly shows a change in its $v$-nullcline. This leads to destruction of the state of equilibrium. In the end, all oscillators come to the vicinity of the equilibrium state. The red oscillator made the spike first, and it turns out to be on the right side of the $v$-nullcline (e), and the blue one, because it was the last one, turns out to be on the left side. Next half-period, they will switch places, and the blue oscillator will be the first to start its spiking behaviour.

Despite the fact that all nullclines almost coincide in projections on a phase portrait, a small deviation is enough to destroy the state of equilibrium. The summand $C_u$ affects the nullcline $\dot{u}=0$. It shifts it up or down depending on the sign. This affects the coordinates of the equilibrium state on the phase plane. The summand $C_v$ shifts the isocline $\dot{v}=0$ left or right. It affects the existence of equilibrium point in general. If $v$-nullcline shifts to the right, then there will be a higher chance of oscillator goes out and starts the spiking behaviour \cite{IZH07}. 

\begin{figure}
\begin{center}
	\resizebox{0.75\columnwidth}{!}{%
  	\includegraphics{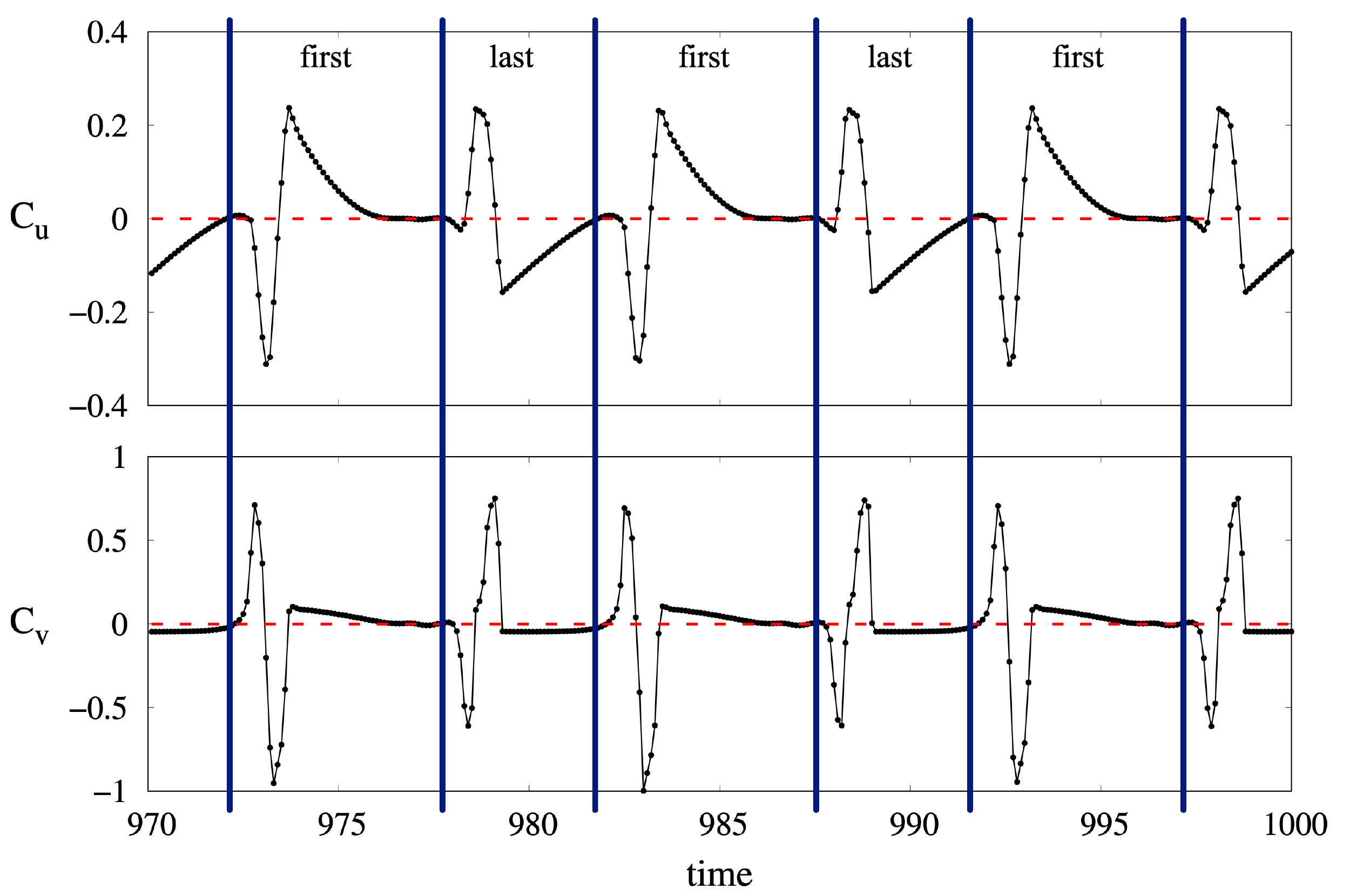} }
\end{center}
\caption{ Coupling terms in first $C_u$ and second $C_v$ equations of the system (\ref{eq:ring_fhn}) prepared for the oscillator $i=60$. Blue vertical lines shows the presence of oscillator near the equilibrium point. Parameters: $\phi=\pi/2+0.1$, $\varepsilon=0.05$, $a=1.001$, $r=0.2$, $\sigma=0.4$, $D=0$.} 
\label{fig:parent_Cuv}       % Give a unique label
\end{figure}

Figure \ref{fig:parent_Cuv} shows temporal implementations of $C_u$ and $C_v$ coupling values. Finding all oscillators near the equilibrium state is accompanied by closeness of both values of $C_u$ and $C_v$ to zero. This happens when the oscillator $i=60$, for which this graph is prepared, starts to spike first and last. Thus, we can assume that the oscillator that is the first one is influenced by the \textit{sign} of these values. From Fig.~\ref{fig:parent_Cuv} we can see that before the phase ``last''  both values of $C_u$ and $C_v$ of the oscillator $i=60$ have very close to zero values. If we consider what happens before the phase ``first'', we can clearly see that the summand $C_v$ has a small negative value. When $C_v$ is less than zero, $v$-nullcline leads to the value $u=-a+|C_v|$, and the nullcline shifts to the right. 

Let us consider what shifts the nullcline. At $\phi=\pi/2+0.1$ the values of the coupling matrix in the second equation (\ref{eq:ring_fhn}) are the following: $b_{vv}=cos(\phi)\approx -0.0999$, $b_{vu}\approx -0.995$. All oscillators near the state of equilibrium have almost identical $v$ values and a small disorder of $u$-variables. Therefore, the main contribution is made by the following summands with cross-link: $\sum_j b_{vu}(u_j-u_i)$. This summand has a negative strength $b_{vu}\approx -0.995$. Correspondingly, if $C_v$ is negative, all $u$-variables of neighbouring oscillators $u_j$ must be greater than $u_i$. This occurs for the oscillator, which comes to the state of equilibrium last. Therefore, the same oscillator becomes the first one in the next half-period. 

Thus, the existence of CR chimeras is mainly caused by a special type of spatial profile. It appears in the system without noise due to the presence of cross-link in the coupling. In the case of a noisy system, there is a slight shift in the parameters at which this profile appears. The presence of an incoherent part is caused by noise exposure. The coupling leads to a special movement of nullclines, and due to the noise several oscillators have the opportunity to start moving first. At $\phi=\pi/2+0.1$ the values of the coupling matrix are as follows: $b_{uu}=b_{vv}=cos(\phi)\approx -0.1$, $b_{uv}=sin(\phi)\approx 0.995$, $b_{vu}=-sin(\phi)\approx -0.995$. In the case of CR-chimeras $\phi=\pi/2-0.1$ the values of cross-link remain the same, and $b_{uu}=b_{vv}\approx 0.1$ changes the sign to the opposite. 

It may give the false impression that the CR chimera can only appear when the impact of cross-linking is much larger than the impact of self-linking. As $\phi$ is an argument of trigonometric functions, such situation should take place when $\phi$ is close to $\pi/2$ and $3\pi/2$. However, the chimera near $3\pi/2$ has not been found yet. Apparently, the positive sign of cross-linking parameters is important, which is negative when $\phi$ is about $3\pi/2$.

\section{Chimera state of type 2}\label{sec:Chimera2}

\begin{figure}
\begin{center}
	\resizebox{0.75\columnwidth}{!}{%
  	\includegraphics{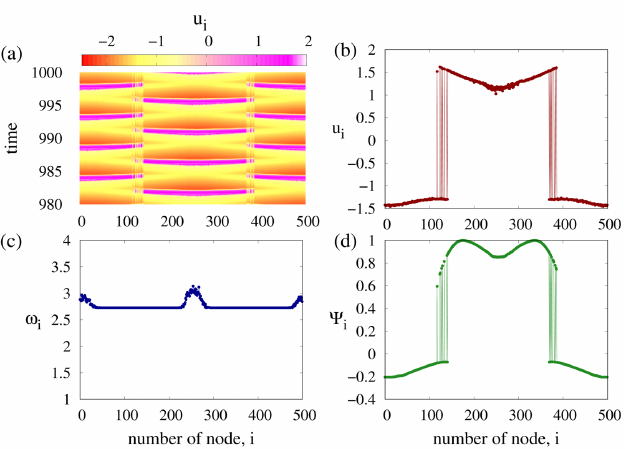} }
\end{center}
\caption{New chimera regime with corresponding space-time plot for the variable $u^t_i$ (a), snapshot (b) at time $t=995.8$, mean phase velocities $\omega_i$ (c) and cross-correlation function (d). Parameters: $\phi=5.4$, $\varepsilon=0.05$, $a=1.001$, $r=0.2$, $\sigma=0.4$, $D=0.0004$. Initial conditions are random distributed in the ring of radius 2.} 
\label{fig:Chimera2}       % Give a unique label
\end{figure}

This section shows that chimera state in the system under study occurs not only when $\phi$ is close to $\pi/2$, but also for other values. The parameter $\phi$ was changed in the interval $\phi\in [0; 2\pi]$. It is found that at value $\phi=5.4$ the system can demonstrate chimera states too, but they qualitatively differ from those described in section \ref{sec:CR_chimera}. The new regime is given in Fig.~\ref{fig:Chimera2}.

Figure \ref{fig:Chimera2},b shows the instantaneous spatial profile at time $t=995.8$.  Two breaks of spatial profile are clearly seen between top and bottom parts. The areas of spatial incoherence born near the breaks. Such an instantaneous images are typical for chimera states appearing in rings of nonlocally connected chaotic systems. For example, logistic maps, Henon maps, R{\"o}ssler systems and many others \cite{OME13, SEM15, SHE18, CHA19}. Such a type of a spatial profile in some literature is called ``phase chimera''. However, in the ensembles of excitable systems (for example, FitzHugh--Nagumo model) this has not been encountered before. 

Position of incoherent and coherent domains does not change in time, and all oscillators continue to demonstrate spiking behaviour (see the space-time plot in Fig.~\ref{fig:Chimera2},a). Due to the stationary position of the incoherent and coherent domains, it is possible to calculate mean phase velocity $\omega_i$ and cross-correlation function $\Psi_{k,i}$ for this state. Figure \ref{fig:Chimera2},c shows mean phase velocities for each oscillator calculated as $\omega_i=2\pi M_i/\Delta T$, $i=1,\dots, N$, where $M_i$ is the number of complete rotations around the origin 
performed by the $i$th unit during the time interval $\Delta T=10^4$. Almost all values of $\omega_i$ lie on a continuous curve. It means that all the oscillators makes their spikes with the same periodicity. Small deviations from the constant value are observed in the minima and maxima of the spatial profile, which indicates the possible dynamics with weak chaos in these regions. This feature will be discussed a while later.

It may seem that each oscillator from the incoherent domain belong to top or bottom parts of spatial profile all the time. However, this is not true. At long time intervals oscillators inside incoherent parts can change their belonging. The oscillators located near the boundaries between incoherent and coherent regions are especially susceptible to this. It is very hard to see it on the space-time plots. Since the incoherent domains are now stationary, we can use the cross-correlation function \cite{BOG16}, which shows temporal correlation between two oscillators (Fig.~\ref{fig:Chimera2},d):
\begin{equation}\label{eq:cross_corr}
\Psi_{k,i}=\frac{\langle \tilde{x}_k(t) \tilde{x}_i (t)\rangle}{\sqrt{\langle \tilde{x}^2_k(t)\rangle \langle\tilde{x}^2_i (t)\rangle}},
\end{equation}
where $\tilde{x}_i (t)=x(t)-\langle\tilde{x}_i (t)\rangle$ is a deviation from the mean value. The brackets $\langle\cdots\rangle$ denote the time averaging. This characteristic is equal to 1 and -1 for in-phase and anti-phase oscillations respectively and is not equal to it for non-synchronized dynamics. 

The calculations of the cross-correlation function in Fig.~\ref{fig:Chimera2},d is prepared in relation to the oscillator  $k=175$ belonging to the top coherent region (Fig.~\ref{fig:Chimera2},b). The cross-correlation function is close to the value 1 only when comparing oscillators from one coherent part. Oscillators from the opposite coherent part are not in anti-phase with them, because the corresponding values of $\Psi$ are not equal to -1. This can be explained by the fact that each oscillator continues to demonstrate spiking behaviour, in which it is impossible to say about the presence of phase or antiphase synchronization, they are just shifted in time. However, Fig.~\ref{fig:Chimera2},d clearly shows that the cross correlation decreases near the boundaries between coherent and incoherent domains. This confirms that some of the oscillators there may be thrown between what a part of the profile they belong to. 

Another interesting feature, which is clearly indicated by the cross-correlation function, is that there is some desynchronization in the middle of one of the coherent domain. There is a local minima at $i\in(200;300)$ in Fig.~\ref{fig:Chimera2},d. This is also one of the features of chimeras arising in ensembles of chaotic oscillators, where this type of chimeras is called amplitude ones. As usual, they have a finite life time in the center of coherent areas. 

The chimera state shown in Fig.~\ref{fig:Chimera2} is obtained for the same parameter values as CR chimera but when the parameter $\phi$ is in a quite large range around the value $5.4$. At this value $\phi$ the coupling matrix has values: $b_{uu}=b_{vv}\approx0.6347$, $b_{uv}=-b_{vu}\approx-0.7727$. Cross-linking and self-linking have a commensurate contribution in this regime in contrast to CR chimera. 

\begin{figure}
\begin{center}
	\resizebox{1\columnwidth}{!}{%
  	\includegraphics{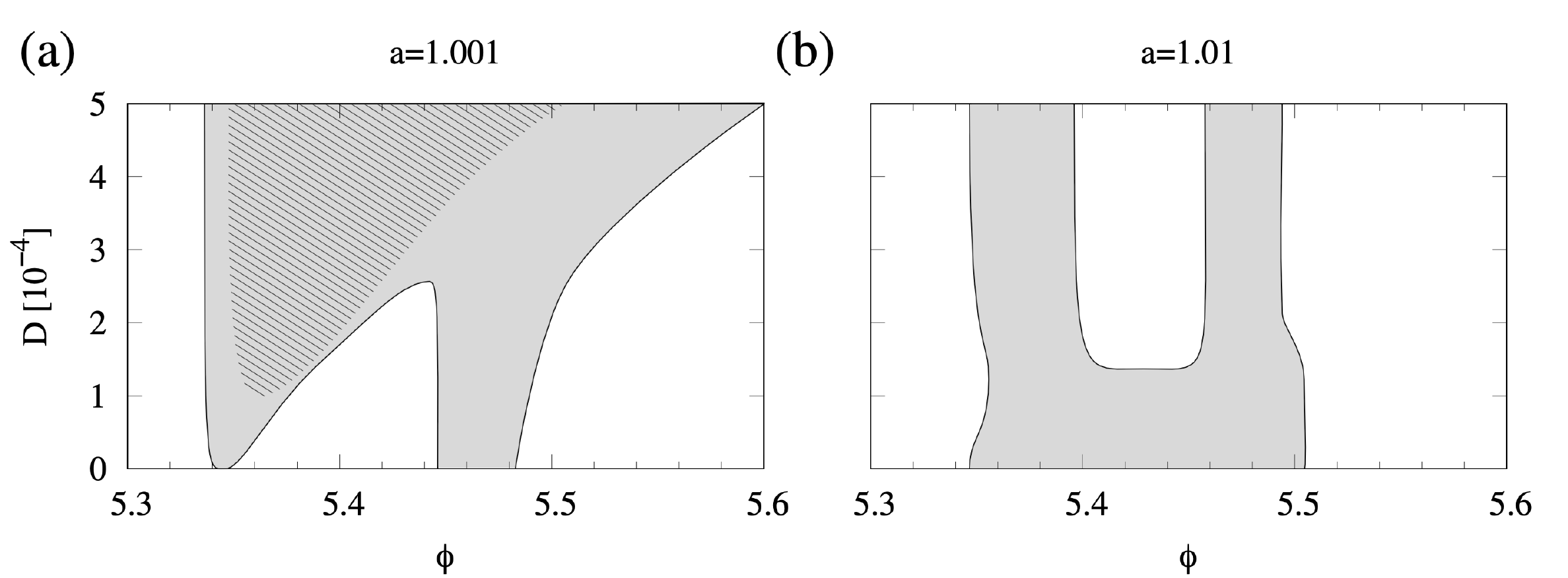} }
\end{center}
\caption{New chimera on the parameter plane ($\phi,D$) shown by gray filled area. The hatched region corresponds to stable chimera state obtained from each random initial condition. Parameters: $\phi=5.4$, $\varepsilon=0.05$, $r=0.2$, $\sigma=0.4$. Initial conditions are random distributed in the ring of radius 2.} 
\label{fig:Chimera2_phi-D}       % Give a unique label
\end{figure}

Let us consider in detail the impact of noise and the $\phi$ parameter on the new chimera state. Fig.~\ref{fig:Chimera2_phi-D} shows the areas of existence of the new chimera on the parameter plane ($\phi,D$) for two values of the parameter $a=1.001$ (near an Andronov--Hopf bifurcation) and $a=1.01$ (far from it). The panel (a) shows the chimera at $a=1.001$. In this mode, the chimera most often coexists with weak oscillations near the equilibrium state. Random initial conditions can lead to either one, or another regime. In such bistable case the chimera can have a finite life time. However, there is a hatched region in Fig.~\ref{fig:Chimera2_phi-D},a where all considered random initial conditions bring up to the chimera. At the same time, the regime is saved even at the integration time of $t=10^6$ with the integration step of $h=0.001$. Experiments with further time increase were not carried out.

Figure \ref{fig:Chimera2_phi-D},b shows the area of existence of chimera states in the plane ($\phi,D$) at $a=1.01$. As for panel (a), several random initial conditions are considered. The chimera coexists with the equilibrium state during the whole chimera region. The clear area of stability, as it was for $a=1.001$, is not found out.

In both cases the most often regime at zero value of noise intensity $D=0$ consists in weak oscillations near the equilibrium point. However there are rare cases when from random initial conditions the spatial profile with the breaks appears (as in Fig.~\ref{fig:Chimera2}, but without incoherent domains). At $D=0$ this regime has a finite lifetime. The chimera can also be obtained, but only from specially prepared initial conditions, and it has a finite lifetime too.

Another difference between the effect of noise at a=1.001 and a=1.01 is the following. The increase of the noise near the bifurcation (Fig.~\ref{fig:Chimera2_phi-D},b at $a=1.001$) leads to an increase in the chimeras range of existence. It is accompanied by the appearance of clear region of stable chimera state obtained from all considered random initial conditions. Noise makes the chimera more stable. In the case of $a=1.01$, the opposite is happening. Noise leads to the decrease and change of the area of existence.

\begin{figure}
\begin{center}
	\resizebox{0.6\columnwidth}{!}{%
  	\includegraphics{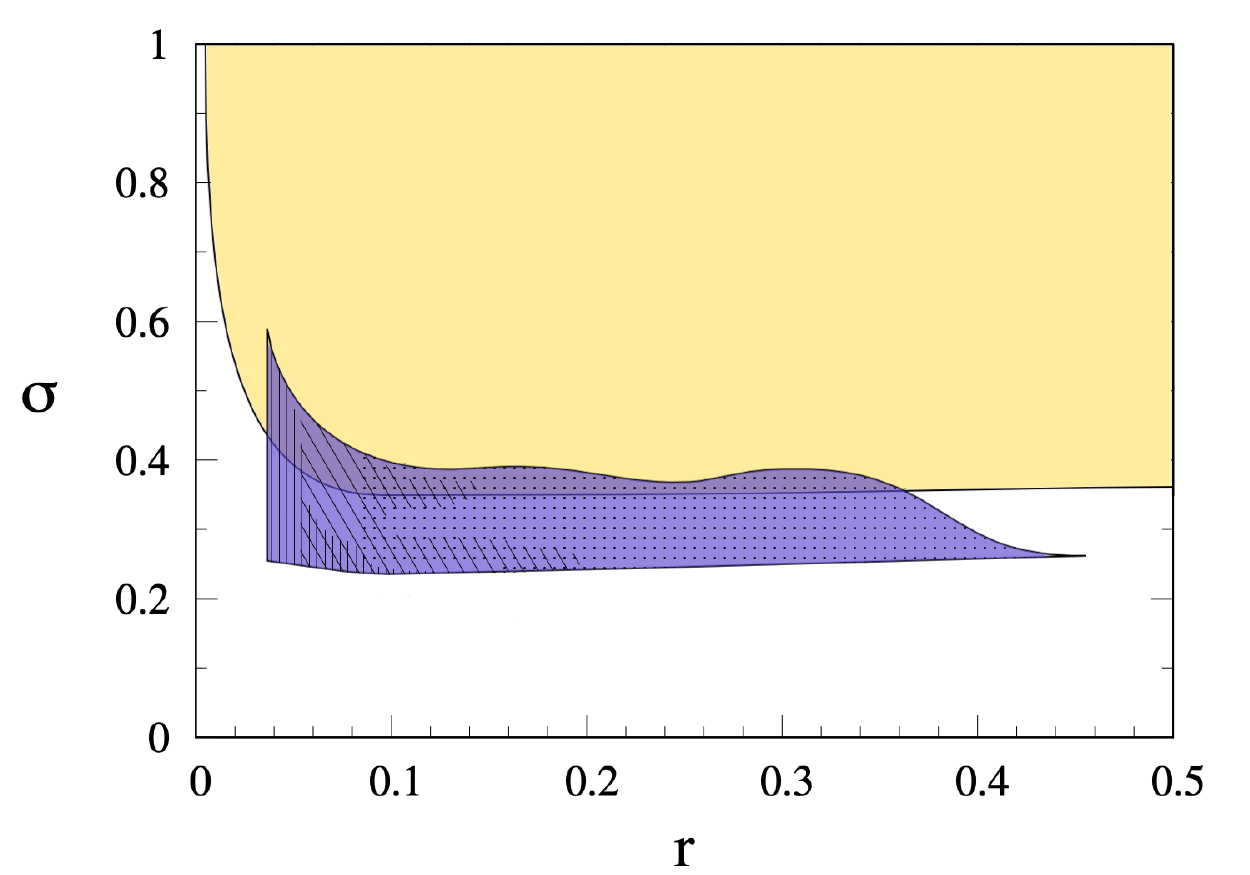} }
\end{center}
\caption{New chimera on the parameter plane ($\sigma,r$) shown by purple area. There are three indicated spatial wave numbers: $K=3$ (vertically hatched), $K=2$ (diagonally hatched) and $K=1$ (dotted). The yellow region correspond to the weak oscillations near the equilibrium point.Parameters: $\phi=5.4$, $\varepsilon=0.05$, $a=1.001$, $D=0.0004$. Initial conditions are random distributed in the ring of radius 2.} 
\label{fig:Chimera2_sigma-r}       % Give a unique label
\end{figure}

The number of incoherent domains in chimera state can be increased by reducing the coupling radius. In the case of chimera shown in Fig.~\ref{fig:Chimera2} it is accompanied by the increase of the spatial wave number. Figure~\ref{fig:Chimera2_sigma-r} shows main regimes obtained on the plane of coupling parameters ($\sigma,r$). It can be seen from the figure that at a large coupling strength $\sigma$ and a large coupling radius $r$, weak oscillations near the state of equilibrium (yellow region) are observed in the system. The same effect has been found out for CR chimera \cite{SEM16}. At small coupling impact the spatial incoherence and various unstable modes are observed in the ensemble (remains white in Fig.~\ref{fig:Chimera2_sigma-r}). Chimera states are realized between these two main areas (purple area in Fig.~\ref{fig:Chimera2_sigma-r}). Inside this area different wave numbers can be obtained. Here only three are indicated: $K=1$ (dotted), $K=2$ (diagonal hatching) and $K=3$ (vertical hatching). The map of regimes is made at $\phi=5.4$, $D=0.0004$. These parameter values correspond to the middle of the stable area of the chimera existence in Fig.~\ref{fig:Chimera2_phi-D}. 

\section{Conclusion}
\label{sec:conclu}
The chimera states arising in the ring of nonlocally coupled excitable FitzHugh--Nagumo systems are considered in this article. Such system can demonstrate coherence-resonance chimera for some values of parameters \cite{SEM16}. Previously, it was assumed that the chimera was caused only by noise. However, no answer was found to the question about periodic switching of the incoherent cluster position. It is shown here that the parameter $\phi$, which is responsible for the influence of cross-linking and self-linking inside the coupling terms, leads to a special type of spatial and temporal dynamics even if the noise is switched off. In this case, a special spatial wave is formed. This mode is the predecessor of the CR chimera. It contains several neighbouring oscillators, which begin to make a spiking event first. The other oscillators joins them later because of the coupling. The position of the first spiking oscillators changes periodically in time. Adding noise to the system causes these oscillators to spike less regularly. This creates an incoherent cluster, and this wave regime is transforming into CR chimera. 

In addition, it is found that if the impact of self-linking and cross-linking is comparable ($\phi\approx 5.4$), the system can demonstrate a chimera, which has not yet been found in the ring of FitzHugh--Nagumo systems. The spatial profile is broken into two parts, and incoherent domains are formed near these breaks. This type of chimera is very typical for the rings of chaotic oscillators and maps, but has not yet been detected for excitable systems. It is shown here that the noise near an Andronov--Hopf bifurcation makes this chimera more stable. The article discusses in detail the effect of noise and coupling parameters on this regime.

%\bibliography{ref}

\end{document}